# Identifying User Behavior from Residual Data in Cloud-based Synchronized Apps

**George Grispos**
g.grispos.1@research.gla.ac.uk
University of Glasgow, United Kingdom

**William Bradley Glisson**
bglisson@southalabama.edu
University of South Alabama, USA

**J. Harold Pardue**
hpardue@southalabama.edu
University of South Alabama, USA

**Mike Dickson**
mike.dickson@gmail.com
Police Scotland, United Kingdom

## Abstract[1]

As the distinction between personal and organizational device usage continues to blur, the combination of applications that interact increases the need to investigate potential security issues. Although security and forensic researchers have been able to recover a variety of artifacts, empirical research has not examined a suite of application artifacts from the perspective of high-level pattern identification. This research presents a preliminary investigation into the idea that residual artifacts generated by cloud-based synchronized applications can be used to identify broad user behavior patterns. To accomplish this, the researchers conducted a single-case, pretest-posttest, quasi experiment using a smartphone device and a suite of Google mobile applications. The contribution of this paper is two-fold. First, it provides a proof of concept of the extent to which residual data from cloud-based synchronized applications can be used to broadly identify user behavior patterns from device data patterns. Second, it highlights the need for security controls to prevent and manage information flow between BYOD mobile devices and cloud synchronization services.

## 1. INTRODUCTION

Mobile devices are integrating into increasingly globally transparent business infrastructures. Gartner predicts that, by 2016, 40 percent of the workforce will be mobile and that the majority of them will possess a smartphone (Gartner, 2012). This evolution potentially impacts a range of business strategies that include network security, device and application development, and data management.

Hence, organizations are investigating various ideas to extend existing information technology infrastructures to include mobile devices (Scheepers & Scheepers, 2004). One possible solution is the implementation of a Bring Your Own Device (BYOD) program. BYOD programs potentially offer several benefits that include an increased level of productivity, mobile device procurement and maintenance cost reduction, increased workforce mobility, location flexibility, increased accessibility and longer working hours (Copeland & Crespi, 2012; Scarfo, 2012).

---

In addition to corporate environments, governments are embracing BYOD programs. Forrester Research (2012) found that 59% of smartphones connected to various government networks were personally-owned devices. TrendMicro (2012) reports that security vulnerabilities found in legitimate smartphone applications can make the extraction of personal and corporate data much easier for cybercriminals. Hence, there are growing concerns that the amount of business and personal information collected by mobile applications could lead to increased end-user profiling (Cleff, 2007). These concerns prompted the research question:

*Is it possible to recover residual data from a reset, resynchronized mobile device running a suite of cloud-based synchronized apps to identify the device user's daily behavioral patterns, social activities, and relationships with other individuals?*

The contribution of answering this question is two-fold. First, it provides a proof of concept that data recovered from cloud-based synchronized applications on a reset, resynchronized mobile device can be used to identify user behavior patterns. Second, it highlights the need for security controls to prevent and manage information flow between BYOD mobile devices and cloud synchronization services.

The paper is structured as follows: Section two discusses related work concerning smartphone security and privacy. Section three presents the methodology and summarizes recovered artifacts. Section four discusses the results. Section five draws conclusions and presents future work.

## 2. RELATED WORK

Smartphone privacy and security have attracted considerable attention from various aspects of academia. Researchers have used a number of methods to assess and understand the risks associated with storing personal information on these devices. Warden and Allen (2011) demonstrated how Apple mobile devices collected and stored location information such as mobile cell towers and Wi-Fi access points. Although no personal information was being recorded, Warden and Allen (2011) argued that should an Apple mobile device be stolen and jail-broken, cybercriminals could easily access and identify previous device owner whereabouts.

The security and privacy of Android devices and applications have also come under scrutiny. TaintDroid (Enck, Gilbert, Chun, Cox, Jung, McDaniel, & Sheth, 2010) was developed to dynamically track the flow of private information through third-party applications installed on an Android device. Enck, et al. (2010) tested thirty random applications with the primary objective of analyzing data leakage from both a privacy and a security perspective. Over one-half the tested applications were reported to be transmitting user location and device information to remote services. Similarly, Gibler, et al. (2012) presented AndroidLeaks which performed a static analysis of code to identify data leakage from Android applications. A third of the 24,000 applications tested were found to store and leak private information such as the device location, Wi-Fi data and audio conversations. Although the authors of TaintDroid and AndroidLeaks have highlighted the amount of personal information being stored and leaked by Android applications, neither author's research focused on assimilating collected information to establish data patterns.

The literature indicates that many researchers focus on the impact of mobile privacy and security from the perspective of individuals and not from the perspective of an organization (Glisson & Storer, 2013). Glisson and Storer (2013) did investigate mobile devices from an organizational privacy and security viewpoint. In terms of organizational security, mobile devices which utilize location and cloud synchronization services are at risk for targeted attacks and can introduce the potential for data leakage within an organization (Grispos, Glisson, & Storer, 2013; Keyes, 2013). Keyes (2013) indicates there is the potential for attackers to use location services to determine where the device owner is located at a specific time, to correlate this information with other sources, to establish associates and provide an indication of the kinds of activities performed in specific locations. However, Keyes does not elaborate on how this information can be extracted by attackers or evaluate any methods for performing this attack.

Miller, et. al., (2012) notes that in a BYOD scenario location, and synchronization services can complicate security issues for an organization. Grispos, et al., (2013) highlighted the technical opportunities for accessing data stored on cloud synchronization services, such as Box, via residual data stored on a mobile device. In corporate environments, the literature identifies security issues related to BYOD solutions as originating from a lack of end-user device controls coupled with a blurring of the distinction between personal and work-related data (Glisson & Storer, 2013; Scarfo, 2012).

Harris, et. al., (2013) conducted a survey with college students, who were about to enter the workforce, to determine their attitudes towards BYOD security. The results from the survey indicated that there is a lack of security awareness from the participants towards BYOD. Twenty percent of the participants admitted that they 'root' or 'jailbreak' their mobile device, potentially creating a major security risk for organizations that would accept these devices on their networks (Harris, et al., 2013). Although researchers have highlighted the security risks of mobile applications in a BYOD context, empirical research has not examined a suite of application artifacts from the perspective of device data pattern identification.

## 3. METHODOLOGY

In order to test the research question of this study, two hypotheses were formulated.

**H1:** Residual data can be recovered from a reset, resynchronized smartphone device that is using cloud-based synchronized applications.

**H2:** Residual data recovered from a reset, resynchronized smartphone is sufficient to correlate device data patterns with known device user behavior patterns.

For the purpose of this research, user behavior patterns are broadly defined as an individual's daily behavioral characteristics, social activities and relationships with other individuals using electronic devices.

### 3.1 Experimental Design

The experimental design employed in this study is the Single Case Pretest-Posttest Quasi Experiment (Campbell & Stanley, 1963). The primary characteristic of a quasi-experiment is the lack of randomization. In a single case design, there is one subject (in this case a smartphone device) that undergoes a measurement, a treatment, and a measurement (O1 X O2) or pretest-treatment-posttest.

The smartphone device used in the pretest-posttest quasi experiment is an HTC One X with 3G data services. An HTC Desire was used in the post-hoc experiment to measure residual data captured from a secondary device. Table 1 – Smartphone device features, highlights the notable features of these devices. The two smartphones selected in this experiment were chosen for their operating systems. The Operating System (OS) for the HTC One X was a recent version at the time of the experiment. This created compatibility issues with 'push-button' forensics solutions. Lack of compatibility forces investigators to use more traditional software development tools. The OS for the HTC Desire represents an older version of Android that is compatible with 'push-button' solutions. This allows for an initial assessment of the output from each approach.

It should be noted that the scope of the quasi experiment was limited in the following ways. The HTC One X smartphone was rooted prior to being used. This experiment was conducted in the United Kingdom using Global System of Mobile Communications technology. The experiment focused on a specific version of the Android operating system and specific versions of Google applications. The HTC

One X solely utilized the Android Debug Bridge for data extraction. The password for the resynchronized smartphone is presumed to be known. The password to the account is not the focus or the research. This is due to legislation that requires suspects to provide this password and encryption information like the Regulation of Investigatory Powers Act of 2000 in the United Kingdom (UK Parliament, 2000) and companies retaining ownership and rights to devices. The emphasis on passwords is further diminished when considered in conjunction with individuals commonly reusing small sets of passwords and current research into resolving this information (Das, Bonneau, Caesar, Borisov, & Wang, 2014; Stobert, 2014).

| Feature | HTC One X | HTC Desire |
|---|---|---|
| Operating System | Android 4.0 (Ice Cream) | Android 2.1 (Éclair) |
| RAM | 1 GB | 567 MB |
| Internal Memory | 32 GB Storage | 512 MB ROM |
| Memory Card | No (Virtual SD Card) | Yes (4 GB) |

Table 1. Smartphone Device Features

The applications included in this experiment were official Google applications that are compatible with the Android operating system v4.0. The Google applications selected for inclusion in this experiment are: Google+ (version 4.0.0.46852618); Google Search (version 1.4.1.278776); Google Calendar (version 201305280); Google Tracks (version 2.0.4); Google Maps (version 6.14.4); Google Drive (version 1.2.182.26); Google Keep (version 1.0.79); and Gmail (version 4.3.1).

It is submitted that the phone used in this experiment is representative of all phones of identical make, model, and configuration. The O1 instance of our design serves as the experimental "control" and is referred to as "Image 1". Image 1 contains the forensically recovered residual data prior to the treatment, the pretest. The treatment is a hard-reset of the device, formatting of the memory card, and a re-synch of the device with the cloud-based apps. The posttest is the recovery of residual data after the treatment and is referred to as "Image 2". Consistent with our research question and hypotheses, the posttest examines the casual impact of the treatment (X) on the amount of forensically recoverable data on the device (O1). A post-hoc forensics test was conducted on a secondary device not included in the control (O1). The results of this test are referred to as "Image 3".

### 3.2 Pretest

The following steps describe how the device was tested for residual data prior to treatment. The residual data forensically recovered from the device serves as the control or baseline against which the posttest results were compared.

1. The HTC One X smartphone's boot-loader was unlocked using the steps on the HTC website (2013) and then the device was 'rooted' using a method described on the CNET website (Griffin, 2012).

2. A desktop computer was used to create a Google account.

3. After the Google account was created, Google contacts were accessed through a desktop web browser to store information for fifteen individuals. The contact information included: first name, last name, mobile phone number and email address.

4. The smartphone device was powered on and the Google account was used to sign-on to Google Services during the initial device setup. An automatic sync with the Google Cloud and the option to allow Google to use location services were also selected during the setup. The device was then configured to use the 3G data services to gain access to the Internet.

5. After the device setup was completed, the Google applications were downloaded and installed using the default installation parameters from the Google Play market.

6. The applications were executed and the test account was used to sign-in to various Google services.

7. Applications were used over a two-week period. Table 2 – Daily Activities presents the activities performed and when they were specifically repeated for each application. Table 3 – Other Activities defines application activities performed on varied days and at varied times. A total of 212 activities were performed using the device which included 58 activities from Table 2 – Daily Activities and 154 activities from Table 3 – Other Activities.

| Occurrence | Activity | Activity Count |
|---|---|---|
| Monday – Friday 9am-10:30am | Google+ Check-in: Kelvin Hall Subway – "Off to Work" | 10 |
| Monday – Friday 10am-11:30am | Google Search: "Starbucks Near Me" and Check-in: "Starbucks Coffee" | 20 |
| Monday – Friday 6pm-8pm | Check-in, Google Tracks: Go for a jog along either University Avenue or Kelvin Way (alternative days) to Home | 10 |
| Tuesday or Friday 8pm-9pm | Google Search, Check-in: Chinese or Indian Restaurant | 4 |
| Monday – Friday 11am-12pm | Google Search: "What's the weather like tomorrow?" | 10 |
| Saturday and Sunday 9am-11am | Google+ Check-in: "At home" | 4 |
| | **Total Daily Activities** | **58** |

**Table 2. Daily Activities**

8. Upon completion of the two week period, artifacts associated with the suite of Google applications were extracted using Android Debug Bridge (ADB). USB debugging was activated on the device and the ADB was used to access the shell command prompt. The Android OS traditionally stores application-related artifacts in the /data/data folder on the User partition (Hoog, 2011). The contents of this folder were copied to a folder on the virtual memory card using the ADB. The virtual memory card was accessed using a write blocker via a desktop computer and a forensic image (Image 1) was created using FTK Imager (AccessData, 2008).

### 3.3 Treatment

The treatment involved the three-step process described below.

1. The HTC One X device was hard-reset and the virtual memory card formatted.

2. The HTC One X device was powered on and the artifact collection process, described in Step 8, was repeated to create a copy of the /data/data folder on the device. This step was implemented to verify that data, related to the Google applications, was no longer on the device.

3. The test account was then used to sign-in to Google services and the applications used in the experiment were reinstalled. The HTC One X device and the applications were allowed to synchronize with the Google Cloud.

### 3.4 Post-test

The purpose of the posttest is to measure the degree to which re-synchronizing a wiped reset device (O2) to the cloud-based apps results in recoverable residual data. Image 2 was produced by repeating the process described in Step 8.

### 3.5 Post-hoc Test

A post-hoc test was conducted on O1, independent of the influence of the treatment (X), by introducing a secondary device. The purpose was to test the degree to which it was possible to recover residual data from O1 with a secondary device while the O1 device was still connected to the cloud-based apps.

The secondary device, an HTC Desire smartphone, was used to sign-in to the Google account using Google services and the applications used in the experiment were installed. This device was allowed to synchronize with the data stored in the Google Cloud. After the synchronization was completed, this device was processed using Cellebrite UFED (version 1.8.5.0). The result of this processing was Image 3. All three forensic images were examined using Physical Analyzer (version 3.7.0.352) and AccessData Forensic Toolkit 4.

| Application | Activity | Activity Count |
|---|---|---|
| Google+ | Posted message on wall; posted on friend's wall; friends posted on wall; sent and received messages; check-in locations; joined, viewed Google+ communities; deleted wall posts, check-ins, messages and left communities. | 81 |
| Google+ Hangouts | Conducted a 'Hangout' | 8 |
| Google Search | Performed searches using typed and voice features. | 10 |
| Google Calendar | Added and deleted entries to calendar. | 7 |
| Google Tracks | 'Tracked' jogging activities. | 4 |
| Google Maps | Requested directions to locations; used Navigation feature for travel. | 15 |
| Google Drive | Saved two XLSX spreadsheet files, two PDF files, and two JPEG images to Drive; deleted PDF files after viewing on the device. | 6 |
| Google Keep | Saved and deleted Notes. | 11 |
| Gmail | Sent and received emails. | 12 |
| **Total Other Activities** | | **154** |

Table 3. Other Activities

Artifacts related to the pretest activities performed using the Google applications listed in Table 2 – Daily activities, were all recovered from Image 1 (O1). All of these artifacts contained timestamp information which matched the date and time the activity took place. The artifacts recovered included Google+ check-in information, posts and messages, as well as Google Search results. One hundred and fifteen (74.6%) out of the 154 artifacts related to the activities in Table 3 – Other activities were also recovered from Image 1.

## 4. ARTIFACTS RECOVERED

It is interesting to note that no artifacts related to Google+ Hangouts were recovered from Image 1. In addition, events or files which were deleted during the experiment were also not recovered from the device. A total of 173 (81.6%) out of 212 activities were found on Image 1. In addition, all fifteen contacts stored in the Google Cloud were also recovered.

The posttest results show that after the HTC One X was hard-reset to factory settings and then re-synched with the Google Cloud, a total of 83 (39%) out of 212 activity artifacts were recovered from Image 2

(O2). The artifacts recovered included 35 artifacts (60%) from Table 2 and 48 artifacts (31%) from Table 3. All fifteen contacts stored in the Google Cloud were recovered from Image 2.

In the post-hoc test, artifacts related to the Google applications were recovered from the HTC Desire which was synchronized with the Google account. In total, 84 (39.6%) out of 212 activity artifacts were recovered from Image 3. The artifacts recovered included 36 (62%) out of 58 from Table 2 and 48 (31%) out of 154 from Table 3. Although the number of artifacts recovered from the Desire and the HTC One X after the resynchronization are similar, different artifacts were recovered from each device. For example, no Google Keep artifacts were recovered from the HTC Desire but they were recovered from the second image of the HTC One X. All fifteen contacts stored in the Google Cloud were recovered from Image 3. Table 4 – Activity Artifacts Recovered summarizes the number of artifacts recovered from each device image for the pretest, posttest, and post-hoc test, as well as providing examples of metadata recovered from each application.

| Application (Real Activity Count) | Metadata Recovered | Pretest (Image 1) | Post-test (Image 2) | Post-hoc (Image 3) |
|---|---|---|---|---|
| Google+ Activities (105) | Timestamps; wall posts; comments; check-in and geo-location points; community feeds and conversations; private messages sent and received; message author information. | 88 | 50 | 47 |
| Google+ Hangouts (8) | - | 0 | 0 | 0 |
| Google Search (34) | Timestamps; search terms, Google URL; and number of visits. | 34 | 0 | 0 |
| Google Calendar (7) | Appointment: title, location, start and end time and creation date and time. | 4 | 4 | 4 |
| Google Tracks (14) | Timestamps; geo-location points; journey coordinates; and Keyhole Markup Language files. | 14 | 14 | 13 |
| Google Maps (15) | Destination information requested; longitude and latitude coordinates; and request time. | 12 | 0 | 5 |
| Google Drive (6) | Timestamps; Favorite files; storage service metadata; files viewed on the device and saved for offline viewing. | 6 | 0 | 6 |
| Google Keep (11) | Notes created; creation and last modified times. | 5 | 5 | 0 |
| Gmail (12) | Email body and subject; sent and received email addresses; email sent/received date and time. | 10 | 10 | 9 |
| Total (212) | | 173 | 83 | 84 |

**Table 4. Activity Artifacts Recovered**

# 5. DISCUSSION

The results of this quasi-experiment are discussed from three perspectives: digital forensics; Bring Your Own Device (BYOD); and high-level device data patterns.

## 5.1 Digital Forensics

The Google artifacts recovered from the device can be used to either confirm or refute the events already discovered from other sources or storage media. Therefore, the evidence recovered from the Google applications can be used to validate or refute a portion of the device owner's social behavior. Furthermore, the Google artifacts recovered can also be correlated with physical evidence to link individuals to certain events, for example, using the check-in data recovered from Google+ with CCTV footage from the related area (Carrier & Spafford, 2003).

From an investigative perspective, there is the potential to use the artifacts recovered to develop social relationship profiles of suspects. Voigt, et al. (2013) reported how law enforcement agencies in Germany are using social networking sites such as Facebook and Google+ to locate personal information and social relationship profiles of suspects. This usually involves police officers befriending suspects on the social network using 'fake' accounts and then examining the social life of the person in question (Voigt, et al., 2013).

Alternatively, law enforcement agencies have also used a social network 'crawler' to identify and analyze these relationships (Voigt, et al., 2013). Although these approaches have not been declared illegal, Voigt, et al. state that evidence gathered using these methods may be inadmissible as evidence in court (Voigt, et al., 2013). The Google artifacts recovered from the experiment in this paper could be considered as an alternative source by law enforcement to identify a suspect's social relationship with other individuals. The Google+ check-ins, posts, messages and pictures, as well as Gmail messages and Google Map locations could all be used to provide investigators with a more complete representation of activities.

## 5.2 BYOD

The implementation of BYOD programs in an organization leads to a potential situation where the boundaries become increasingly blurred between personal and corporate data. In a BYOD environment it is plausible that a personally-owned device could be accessing corporate data while interacting with cloud synchronization services (Morrow, 2012). This presents an opportunity for a malicious insider to use these services to steal corporate data and save it in cloud storage services such as Dropbox (Morrow, 2012). This scenario recently resulted in IBM restricting its workforce from using cloud services, as well as Siri, Apple's personal assistant (Leyden, 2012). The results of this experiment further highlight the potential risk that cloud synchronization applications can introduce to an organization in a BYOD context.

The experiment's results demonstrated that application information is synchronized and stored offsite. When the HTC One X was reset to factory settings, the Google applications were reinstalled and synchronized with the Google Cloud. A total of 83 activity artifacts were restored to the device and recovered from Image 2. This represents nearly 48% of the activity artifacts which were recovered from Image 1. Furthermore, this information was only secured by a single username and password. The recent attacks on Google (Fletcher, 2010) and Evernote (Forbes Online, 2013), have highlighted that single-sign-on systems can further complicate BYOD scenarios for corporate organizations.

Another threat identified in the post-hoc test is the potential for an attacker to hijack a specific account without the user being aware they are under attack. When a second device, an HTC Desire was synchronized with the same credentials as those on the HTC One X, the device could be used to access all the information stored in the Google Cloud while it was still accessible from the One X. There was no notification from Google that an additional device was accessing the experimental data set. This could lead to the following scenarios:

- Corporate information could be compromised from a 'piggy-back device'. The organization and device owner may be unaware this has occurred; or
- A victim could use a secondary device primarily used by another individual such as a spouse or family member to access corporate information. The secondary device, if stolen or compromised could expose residual data to attack; or
- A victim could be locked out of his/her account causing the device to no longer synchronize with the Google Cloud.

The results of the experiment coupled with these scenarios highlight challenges associated with the management and protection of corporate data.

### 5.3 Pattern Development

Mobile location-based services are predominantly used to determine where a mobile device user is located. These services are used to not only tell the device user where and how to get to their destination, but also to disclose which friends are nearby, what the weather forecast is and what places of interest are located nearby (Vaughan-Nichols, 2009). The problem arises when this location information is integrated with personal or business information.

Google currently requires users to sign-in to their Google accounts to use any smartphone application, Gmail and any other Google service (Bauer, Bravo-Lillo, Fragkaki, & Melicher, 2013). Google can, potentially, assimilate data about an individual's habits using any of their services with their activities on Google+. This integration of information can be dangerous from a high-level pattern recognition perspective, particularly in corporate environments.

However, the amount of information stored in Google applications is of greater concern when lost and/or stolen mobile devices, such as smartphones, can be used in social engineering attacks against an organization (Friedman & Hoffman, 2008; Landman, 2010; Weippl, Holzinger, & Tjoa, 2006). Should a device which has been used for both work and personal use be lost or stolen, there is the potential for the device to be 'rooted' and the data on the device used for a social engineering attack.

The artifacts recovered from the pretest and posttest images, in relation to the activities in Table 2 – Daily activities and Table 3 – Other activities, indicate that this data can be used to establish high-level device data patterns. All of the artifacts from Table 2 and 74% of the artifacts from Table 3 were recovered from the pretest image, while 60% of the artifacts from Table 2 and 31% of the artifacts from Table 3 were recovered from the posttest image. The recovery and clustering of timestamps for the activities presented in Table 2 suggests that it may be possible to identify high-level blocks of time when an individual is typically engaged in some activity. This type of information can be valuable to a social engineering attacker who would like to know when the device owner may be away from his/her workstation or office. The results from Image 2 (posttest) also indicate that a substantial portion of this data is being stored in the cloud. The synchronization of a device, with no personal data, with the Google Cloud retrieved nearly 48% of the activity artifacts which were recovered from Image 1.

### 6. CONCLUSIONS AND FUTURE WORK

The results of the quasi experiment described in this paper provide preliminary support for both hypothesis 1 and 2. Substantial residual data of known user activities were recovered from the pretest and posttests, 81% and 39% respectively. The post-hoc test also resulted in a recovery of 39% of known user behaviors.

This initial investigation provides a proof of concept that known user behavior can be correlated with high-level device data patterns based on data generated from a smartphone using cloud-based synchronized apps. The clustering of the timestamps for the activities presented in Table 2 indicates high-level data patterns are identifiable. The research also indicates that a substantial portion of this data is being stored in the cloud. The synchronization of a device with no personal data with the Google Cloud retrieved 83 activity artifacts. This represents nearly 48% of the activity artifacts which were recovered from Image 1. This finding reinforces the need to investigate security controls to be able to prevent or manage information flow between BYOD mobile devices and cloud synchronization services. The experiment also highlights a potential hijacking opportunity. A secondary device can be used to login to the Google Cloud and synchronized without the victim being aware or notified that this action has occurred.

This study provides a foundation for expanded, richer, more extensive and real-world based datasets for individuals and organizations. The data raises additional questions about the discrepancies between data extraction methods like the android debug bridge and the Cellebrite extraction tool. These inconsistencies should be examined in future studies.

Future research will investigate the introduction of mobile devices into real-world environments in order to track, visualize and compare algorithms designed to de-couple business and personal data. The idea is not only to be able to look backward at the static residual data on the device to develop detailed device profiles but to be able to investigate effective ways to link individuals to specific device behavior. The ultimate goal is to develop algorithms that can link devices to individuals and predict future behavior with a high degree of certainty. Success in this area could have positive implications in minimizing the current risk associated with BYOD solutions in organizations. Detailed activity profiles created from the algorithms could be used to alert security personnel to a suspicious activity. The establishment of metrics to determine an organizations' comfort level with an employee's mobile device activities could provide insight into potential security issues and, potentially, mitigate BYOD concerns for organizations. In addition, future work will expand the experiment to include a variety of smartphones and Operating Systems (OS). The focus is to evaluate pattern identification and validation across multiple devices and OSs.

## 7. REFERENCES

AccessData. (2008). Forensic Toolkit, from http://www.accessdata.com/

Allan, A., & Warden, P. (2011). Got an iPhone or 3G iPad? Apple is recording your moves, from http://radar.oreilly.com/2011/04/apple-location-tracking.html

Bauer, L., Bravo-Lillo, C., Fragkaki, E., & Melicher, W. (2013). *A comparison of users' perceptions of and willingness to use Google, Facebook, and Google+ single-sign-on functionality.* Paper presented at the 2013 ACM workshop on Digital identity management.

Campbell, D. T., & Stanley, J. C. (1963). *Experimental and quasi-experimental designs for research*: Houghton Mifflin Boston.

Carrier, B., & Spafford, E. H. (2003). Getting physical with the digital investigation process. *International Journal of Digital Evidence, 2*(2), 1-20.

Cleff, E. B. (2007). *Privacy issues in mobile advertising*. Paper presented at the BILETA 22nd Annual Conference.

Copeland, R., & Crespi, N. (2012). *Controlling enterprise context-based session policy and mapping it to mobile broadband policy rules*. Paper presented at the 16th International Conference on Intelligence in Next Generation Networks.

Das, A., Bonneau, J., Caesar, M., Borisov, N., & Wang, X. (2014). *The Tangled Web of Password Reuse*. Paper presented at the Network and Distributed System Security Symposium, San Diego, CA. https://security.cs.princeton.edu/publications/

Enck, W., Gilbert, P., Chun, B.-G., Cox, L. P., Jung, J., McDaniel, P., & Sheth, A. N. (2010). *TaintDroid: an information-flow tracking system for realtime privacy monitoring on smartphones*. Paper presented at the 9th USENIX conference on Operating systems design and implementation.

Fletcher, O. (2010). Google Attack Targeted 'Gaia' Password System, from http://www.pcworld.com/article/194557/article.html

Forbes Online. (2013). Evernote joins the notably hackable club, from http://www.theregister.co.uk/2013/03/04/evernote_password_reset/

Forrester Research. (2012). BYOD in Government: Prepare For The Rising Tide.

Friedman, J., & Hoffman, D. V. (2008). Protecting data on mobile devices: A taxonomy of security threats to mobile computing and review of applicable defenses. *Information, knowledge, systems management, 7*(1), 159-180.

Gartner. (2012). Gartner Says 821 Million Smart Devices Will Be Purchased Worldwide in 2012, from http://www.gartner.com/newsroom/id/2227215

Gibler, C., Crussell, J., Erickson, J., & Chen, H. (2012). Androidleaks: Automatically detecting potential privacy leaks in android applications on a large scale *Trust and Trustworthy Computing* (pp. 291-307): Springer.

Glisson, W. B., & Storer, T. (2013). *Investigating Information Security Risks of Mobile Device Use within Organizations*. Paper presented at the The 19th Americas Conference on Information Systems (AMCIS), Chicago.

Griffin, B. (2012). How to root the HTC One X, from http://reviews.cnet.co.uk/mobile-phones/how-to-root-the-htc-one-x-50010022/

Grispos, G., Glisson, W. B., & Storer, T. (2013). *Using smartphones as a proxy for forensic evidence contained in cloud storage services*. Paper presented at the 46th Hawaii International Conference on System Sciences, Maui, Hawaii.

Harris, M. A., Patten, K., & Regan, E. (2013). *The Need for BYOD Mobile Device Security Awareness and Training*. Paper presented at the The 19th Americas Conference on Information Systems (AMCIS), Chicago.

Hoog, A. (2011). *Android Forensics - Investigation, Analysis and Mobile Security for Google Android* Syngress.

HTC. (2013). Unlock Bootloader, from http://www.htcdev.com/bootloader/faq

Keyes, J. (2013). *Bring Your Own Devices (BYOD) Survival Guide*: CRC Press.

Landman, M. (2010). *Managing smart phone security risks*. Paper presented at the 2010 Information Security Curriculum Development Conference.

Leyden, J. (2012). IBM bans Dropbox, Siri and rival cloud tech at work, from http://www.theregister.co.uk/2012/05/25/ibm_bans_dropbox_siri/

Miller, K. W., Voas, J., & Hurlburt, G. F. (2012). BYOD: Security and Privacy Considerations. *IT Professional, 14*(5), 53-55.

Morrow, B. (2012). BYOD security challenges: control and protect your most sensitive data. *Network Security, 2012*(12), 5-8.

Scarfo, A. (2012). *New Security Perspectives around BYOD*. Paper presented at the Seventh International Conference on Broadband, Wireless Computing, Communication & Applications.

Scheepers, H., & Scheepers, R. (2004). *The implementation of mobile technology in organizations: expanding individual use contexts*. Paper presented at the 25th International Conference on Information Systems, Washington DC.

Stobert, E. (2014). *The agony of passwords: can we learn from user coping strategies?* Paper presented at the CHI '14 Extended Abstracts on Human Factors in Computing Systems, Toronto, Ontario, Canada.

TrendMicro. (2012). *Security in the Age of Mobility*.

UK Parliament. (2000, December 8, 2000). Regulation of Investigatory Powers Act 2000 Retrieved May 7, 2014, from http://www.legislation.gov.uk/ukpga/2000/23/contents

Vaughan-Nichols, S. J. (2009). Will Mobile Computing's Future Be Location, Location, Location? *Computer, 42*(2), 14-17. doi: 10.1109/MC.2009.65

Voigt, S., Hinz, O., & Jansen, N. (2013). *Law enforcement 2.0: the potential and the (legal) restrictions of Facebook data for police tracing and investigation*. Paper presented at the 21st European Conference on Information Systems, Utrecht.

Weippl, E., Holzinger, A., & Tjoa, A. M. (2006). Security aspects of ubiquitous computing in health care. *Elektrotechnik und Informationstechnik, 123*(4), 156-161.